\pdfoutput=1

\documentclass[11pt]{article}

\usepackage[]{acl}

\usepackage{times}
\usepackage{latexsym}
\usepackage{graphicx}
\usepackage{multirow} 
\usepackage{supertabular}
\usepackage{caption}
\usepackage{adjustbox}
\usepackage{stfloats}
\usepackage[T1]{fontenc}

\usepackage[utf8]{inputenc}
\usepackage{xspace}
\usepackage{hyperref}
\usepackage{microtype}

\usepackage{tabularx}

\definecolor{ultramarine}{RGB}{0,32,96}

\newcommand{\model}{InvestLM\xspace}
\title{\model: A Large Language Model for Investment using Financial Domain Instruction Tuning }

\author{\bf Yi Yang \thanks{~~Correspondence: \texttt{imyiyang@ust.hk}},
        Yixuan Tang ,
        Kar Yan Tam \\ 
        \\
    Hong Kong University of Science and Technology\\
    \texttt{\{imyiyang,yixuantang,kytam\}@ust.hk}\\
    }

\begin{document}
\maketitle
\begin{abstract}
We present a new financial domain large language model, \model, tuned on LLaMA-65B \citep{llama}, using a carefully curated instruction dataset related to financial investment. Inspired by less-is-more-for-alignment \citep{zhou2023lima}, we manually curate a small yet diverse instruction dataset, covering a wide range of financial related topics, from Chartered Financial Analyst (CFA) exam questions to SEC filings to Stackexchange quantitative finance discussions. \model shows strong capabilities in understanding financial text and provides helpful responses to investment related questions. Financial experts, including hedge fund managers and research analysts, rate  \model's response as comparable to those of state-of-the-art commercial models (GPT-3.5, GPT-4 and Claude-2). Zero-shot evaluation on a set of financial NLP benchmarks demonstrates strong generalizability.  From a research perspective, this work suggests that a high-quality domain specific LLM can be tuned using a small set of carefully curated instructions on a well-trained foundation model, which is consistent with the Superficial Alignment Hypothesis \citep{zhou2023lima}. From a practical perspective, this work develops a state-of-the-art financial domain LLM with superior capability in understanding  financial texts and providing helpful investment advice, potentially enhancing the work efficiency of financial professionals. We release the model parameters to the research community\footnote{\model adopts the same licensing terms as LLaMA \citep{llama}. Link: \url{https://github.com/AbaciNLP/InvestLM}}.
\end{abstract}

\section{Introduction}
Large language models (LLMs) have significantly changed the paradigm of natural language processing \citep{brown2020language,llama} and hold great potential for artificial general intelligence \citep{bubeck2023sparks}. Several financial domain LLMs have also been developed with the hope of processing massive financial texts and enhancing investment and financial decision-making for investors and financial professionals.

However, three challenges may hinder the broad development and adoption of financial domain LLMs.  First, BloombergGPT \citep{wu2023bloomberggpt}, a foundation model with 50 billion parameters trained on Bloomberg's proprietary data, is not publicly available. Thus, the community cannot study its capabilities in financial tasks. Second, while other commercialized LLMs such as ChatGPT and Claude-2 are accessible via API, their model parameters are not publicly available either, making it expensive to investigate their financial task capability. Third, the research community has released several LLMs fine-tuned on financial NLP tasks, such as FinMA \citep{xie2023pixiu} and FinGPT \citep{yang2023fingpt}. 
However, these models are not only smaller in size \footnote{At the time of writing, the largest FinMA model is 30B and FinGPT is 7B.} but also exhibit poor performance when generalizing to financial NLP tasks beyond their instructed tasks.

\begin{table*}[]
\small
\centering
\begin{adjustbox}{width={\textwidth},totalheight={0.96\textheight},keepaspectratio}%
\begin{tabularx}{\textwidth}{p{0.08\textwidth}|p{0.88\textwidth}}
\hline
\multicolumn{2}{p{0.99\textwidth}}{ \textcolor{ultramarine}{The railroad company saw earnings roll a bit off track when it reported July 20. The company slightly missed analysts’ revenue expectations and reported earnings per share (EPS) that matched Wall Street’s estimates. That EPS performance might be fine for many companies, but it rang warning bells for CSX investors because it was the first time in five years that CSX failed to surpass EPS estimates. The company saw declining volume in a number of key products it transports, including agricultural and food, chemicals, and forest. Volume growth in coal provided some locomotion. Intermodal, which means using two transport modes to move freight (which often means converting freight from highway to intermodal rail), continued to struggle in the latest quarter. Railroads have grappled with worker shortages and supply chain issues. The latest CSX earnings raises questions about overall U.S. economic health. }}
\\
\multicolumn{2}{p{\linewidth}}{} 
\\
\multicolumn{2}{p{\linewidth}}{ \textcolor{ultramarine}{Given the recent performance of the railroad company, is it advisable to buy, hold, or sell stocks in the sector?} }\\
\hline
\textbf{\model} 
& Based on the information provided, the stock performance appears to be mixed. While revenue missed expectations, EPS met estimates, and coal volume provided some growth. On the other hand, declining volume in key product categories along with ongoing labor shortages and supply chain issues raise concerns. Overall, the outlook seems uncertain. As a result, a "hold" recommendation may be more appropriate than an outright "buy" or "sell" at this time. Investors should continue to monitor the economic indicators and company fundamentals before making an investment decision. \\
&\\
\hline

\textbf{GPT-3.5} & I'm an AI language model, and I can provide you with some information to consider, but please note that investing decisions should be based on thorough research and analysis. The recent performance of the railroad company, CSX, seems to have raised concerns about its earnings and volume decline in key product categories. The struggles with intermodal transportation and overall supply chain issues may also impact the company's performance.
 \\
 & When making investment decisions, it's important to consider factors such as the company's long-term prospects, industry trends, and the overall market conditions. Here are a few points to keep in mind: \\
 
 & 1. Research the sector: Look into the railroad industry as a whole and analyze its growth potential, overall trends, and challenges. Understand how other companies within the sector are performing.
 2. Evaluate the company's fundamentals: Assess CSX's financial statements, including revenue trends, profit margins, and debt levels. Consider the company's competitive position, management team, and their strategies for addressing the challenges they face.
 3. Consider market conditions: Examine the wider economic environment and its impact on the railroad industry. Factors such as macroeconomic trends, government regulations, and consumer demand can influence railroad stocks.
 4. Diversification: If you already have holdings in the sector, consider the diversification of your portfolio. Adding more railroad stocks may increase exposure to similar challenges. Diversifying across other sectors can help mitigate risk.
 \\
 & Ultimately, it is advisable to consult with a financial advisor or conduct further research to make an informed decision about buying, holding, or selling stocks in the railroad sector. They can provide personalized advice based on your financial goals and risk tolerance.\\
 \\
\hline

\textbf{GPT-4} & As an AI language model, I can provide you with a logical analysis based on data provided, but can't provide personalized financial advice. Here are some considerations: \\

& 1. **Buy**: If you believe the shortcomings in the CSX earnings report are temporary and that the company can overcome the challenges it faces, you may choose to buy. A potential advantage is that often when companies or sectors are performing relatively poorly, their stocks could potentially be undervalued, offering a buying opportunity for long-term investors. Furthermore, any recovery in the volume of key products could provide a significant boost to the company's earnings in the future.
\\
& 2. **Hold**: If you already own shares of CSX and believe that the company will weather the difficulties over time, holding could be a viable strategy. If you're optimistic about the long-term prospects of the rail industry or the U.S. economy as a whole, there could be value in waiting for improvements.
\\
& 
3. **Sell**: If you believe the challenges that CSX faces are indicative of persistent issues within the rail industry or the broader economy, it might be a signal to reduce your exposure to this sector.
\\
& Overall, the decision should be based on your risk appetite, your faith in the company's management to turn things around, and your understanding of the overall industry and economic climate. It may be wise to consult with a financial advisor or conduct further stock analysis before making a decision.

\\
\hline
\textbf{Claude-2} & 
Based on the information provided, here is my assessment of whether it would be advisable to buy, hold, or sell stocks in the railroad sector:
\\
& - The miss on revenue expectations and meeting EPS expectations rather than beating them could indicate weaker demand for transportation services. Declining volume in key product categories also points to softer economic activity. This raises questions about the overall economic outlook.
\\
& - Continued struggles in intermodal transportation suggests ongoing supply chain issues are weighing on the sector. Worker shortages are also a challenge railroads are facing. 
\\
&- The recent CSX earnings report was the first time in 5 years they did not beat EPS expectations, suggesting the company is facing more headwinds now compared to the past several years.
\\
&- Taken together, the mixed results from CSX and the issues around demand, volumes, supply chains and labor point to near-term weakness in the railroad sector. While transportation volumes tend to recover as economies strengthen, the current environment raises uncertainties.
\\
& Given this analysis, I would say it is not advisable to buy additional stocks in the railroad sector at this time. The recent performance and underlying issues suggest risk that future earnings and growth may disappoint. For current investors, holding positions makes sense if taking a long-term view, but consider trimming or selling portions of positions to reduce risk exposure until the outlook becomes clearer. In summary, a ""hold"" viewpoint seems most prudent, with a caveat to potentially reduce risk through selective selling.

\\
\hline
\end{tabularx}
\end{adjustbox}
\caption{\model, GPT-3.5, GPT-4, and Claude-2 response to the input question.}
\label{tab:example}
\end{table*}

To address the challenges, in this work, we develop a new financial domain LLM: \model. \model is trained on the LLaMA 65B, using a carefully curated instruction dataset related to finance and investment. We release the parameters of \model and adopt the same licensing terms as LLaMA. We evaluate the \model's utility in providing helpful investment advice by collaborating with financial experts, including hedge fund managers and research analysts. Experts rate \model's responses as comparable to  those of state-of-the-art commercial models (GPT-3.5, GPT-4 and Claude-2). An example of \model and three commercial models' responses are presented in Table \ref{tab:example}. According to expert feedback, GPTs often give vague responses to investment-related questions, possibly because of its safety requirements. In scenarios where commercial models offer decisive answers, \model typically arrives at a logical investment conclusion similar to these commercial models, but it is often more concise. Case studies also reveal that \model effectively mitigates the hallucination problems often found in LLaMA's responses. In addition,  \model shows strong generalizability on several financial NLP tasks - those that are not used in the instruction tuning. 

In addition to contributing a new financial domain LLM, we are also interested in understanding the behaviors of domain instruction tuning.  We find that a diverse set of domain instructions is very effective in "transforming" a high-quality foundation model (such as LLaMA-65B) into a high-quality domain-specific model -  suggesting the consistency with the Superficial Alignment Hypothesis \citep{zhou2023lima}.
Second, we discover that generic instructions, like those used in Alpaca \cite{alpaca}, can detrimentally impact the performance of instruction-tuned models on domain tasks. This underscores the importance of curating domain-specific instructions. Together, these findings provide insights into how to fine-tune a foundation model for a specific domain.

The rest of the paper is organized as follows. First, we describe our manually curated financial domain instruction dataset. This is followed by an overview of the instruction tuning procedures for developing \model. Subsequently, we present expert evaluation results and report the performance of \model, comparing it with other LLMs on a set of financial NLP benchmarks. Lastly, we conduct studies to better understand the behaviors of domain instruction tuning.

\section{Instruction Dataset}
Our goal is to construct a financial LLM capable of understanding financial text and providing helpful responses to investment-related questions. The Superficial Alignment Hypothesis \citep{zhou2023lima} posits that \textit{A model’s knowledge and capabilities are learned
almost entirely during pretraining, while alignment teaches it which subdistribution of formats should
be used when interacting with users}. Thus, we hypothesize that a well-trained large language model, such as LLaMA-65B, has already acquired the ability to understand financial text, given that its training corpus might encompass financial content. So, to enable LLaMA to interact with investment-related questions, we should fine tune it with specifically crafted instructions. 

We manually curate an instruction dataset from the following resources: 

\paragraph{Stackexchange QFin.} We select a set of questions along with their corresponding high-vote answers from the quantitative finance board.

\paragraph{CFA questions.} We choose a set of questions with detailed answers (including explanations) from the Chartered Financial Analyst (CFA) exams.

\paragraph{Academic Journals.} We choose articles from top financial economics journals (such as \textit{Journal of Finance}) and manually create questions related to asset pricing and risk management, and solicit answers from the articles. 

\paragraph{Textbooks.} We choose several classic finance textbooks (such as \textit{Investments, 10th Edition}), and select the exercises from these textbooks as instruction/inputs and find their corresponding standard answers as outputs. 

\paragraph{SEC filings.} We select earnings conference call transcripts and SEC filings, curate several questions based on the content of the filings, and then extract the corresponding answers from the text.

\paragraph{Financial NLP tasks.} We reformat several financial NLP tasks, such as financial sentiment analysis, and numerical reasoning as instructions. It's worth noting that, even for instructions derived from financial NLP tasks, we do not directly use the corresponding label as the output. Instead, we manually provide a response to augment the label.

\paragraph{Investment questions.} We brainstorm financial and investment-related questions, then use ChatGPT and Claude-2 to generate answers. Subsequently, we manually verify and paraphrase the best answer to use as the output.

For each instruction resource, we gather several hundred examples, with each example being a tuple comprising instruction, input, and output. The detailed breakdown of our financial domain instruction dataset is shown in Table \ref{tab:data}. Compared to other general-purpose instruction datasets, such as that used in Alpaca \citep{alpaca}, our finance-specific instruction dataset has much longer instruction/input and output lengths. Most importantly, our dataset is manually curated to encompass topics related to financial investment.

\begin{table}[]
\small
\begin{tabular}{llll}
\hline
 & Size & Input len.  & Resp. len.\\ \hline
All & 1,335 & 152.9 & 145.5 \\ \hline
Stackexchange & 205 & 19.4 & 296.2 \\
CFA & 329 & 125.6 & 157.4 \\
Academic Journals & 200 & 169.3 & 74.8 \\
Textbooks & 200 & 128.9 & 136.6 \\
SEC Filings & 80 & 316.2 & 88.2 \\
Financial NLP tasks & 200 & 325.9 & 74.5 \\
Investments & 119 & 72.7 & 144.3 \\ \hline
\end{tabular}
\caption{Detailed breakdown of our financial domain instruction dataset. Input len. and Resp. len.  denotes the average number of tokens of instruction/input and output response respectively.}
\label{tab:data}
\end{table}

\section{Instruction Tuning on LLaMA}
We train \model on LLama-65B using our financial domain instruction dataset \cite{llama}. Specifically, we use the Low-rank adaptation (LoRa) method \cite{hu2021lora} to tune the model parameters in order to enhance the training efficiency. We set the rank to 16. We specifically target LoRa modules as "q\_proj," "k\_proj," "v\_proj," "down\_proj," "gate\_proj," and "up\_proj". 

Moreover, most financial texts such as SEC filings, corporate disclosures, and analyst reports are long text, usually consisting of thousands of tokens. Thus it is important that the financial domain LLM can handle long text.  To this end, we utilize Linear Rope Scaling \cite{chen2023extending} on a scale of 4 to increase context size to \textbf{8,192}, which improves \model's ability to handle long financial texts. The detailed training prompt format is shown in Appendix \ref{sec:training_format}.

Following standard fine-tuning hyperparameters, we train the LLaMA-65B model for 15 epochs. The learning rate is set to 3e-4, and the batch size to 16 examples. In the subsequent analysis, we also instruction tune a LLaMA-7B model. For this model, we conduct training for 12 epochs with a learning rate of 3e-3 and a batch size of 32 samples. For simplicity, we denote \model-65B as \model, and LLaMA-65B as LLaMA.

\section{Expert Evaluation}
The primary goal of this work is to build a financial LLM capable of understanding financial text and providing helpful responses to investment related questions. To test whether \model can offer helpful responses, we collaborate and conduct interviews with a group of six financial experts, including hedge fund managers and financial analysts. 

\noindent\textbf{Baselines}. We compare \model with three state-of-the-art commercial models, \textbf{GPT-3.5}, \textbf{GPT-4} and \textbf{Claude-2}. OpenAI's GPT-3.5 and GPT-4 are large language models tuned with reinforcement learning from
human feedback (RLHF) \citep{ouyang2022training}. Anthropic's  Claude-2 is a large language model that can take up to 100K tokens in the user's prompt. \footnote{\url{https://www.anthropic.com/index/claude-2}} Responses from all
baselines are sampled throughout August 2023.

We manually write 30 test questions that are related to financial markets and investment. For each question, we generate a single response from  \model and the three commercial models.  We then ask the financial experts to compare \model responses to each of the baselines and label which response is better or whether neither response is significantly better than the other.

In addition to the expert evaluation, we also conduct a GPT-4 evaluation, following the same protocol used in \citep{zhou2023lima}. Specifically, we send GPT-4 with exactly the same instructions and data annotations, and ask GPT-4 which response is better or whether neither response is significantly better than the other. The expert evaluation interface and GPT-4 evaluation prompt are presented in Appendix \ref{sec:interface}. 

\begin{figure*}[htbp]
    \centering
    \begin{minipage}{0.48\textwidth}
        \centering
        \includegraphics[width=1\textwidth]{expert_eval} 
        \caption{Expert evaluation.}
        \label{fig:expert_eval}
    \end{minipage}\hfill
    \begin{minipage}{0.48\textwidth}
        \centering
        \includegraphics[width=1\textwidth]{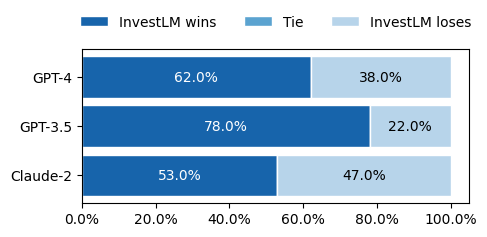} 
        \caption{ GPT-4 evaluation.}
        \label{fig:gpt4_eval}
    \end{minipage}
\end{figure*}

The expert evaluation and GPT-4 evaluation results are presented in Figure \ref{fig:expert_eval} and Figure \ref{fig:gpt4_eval}. These results indicate that financial experts rate \model's responses as either comparable to or better than those of the GPT-3.5 and GPT-4 models. This expert assessment aligns with GPT-4's own evaluation, which also prefers \model's responses. While financial experts tend to favor Claude-2's responses over \model most of the time, GPT-4 shows a preference for \model's responses. Overall, it is encouraging to observe that our domain-specific instruction tuning effectively generates helpful answers to investment-related questions, especially considering that its foundational model, LLaMA, frequently produces hallucinations, as shown in Section \ref{sec:investLM vs. LLaMA}.

\noindent\textbf{Inter-Annotator Agreement}. We compute the Inter-Annotator Agreement of evaluation by the following rules.  We compute the tie-discounted accuracy for each example between the majority-vote label and each annotator's label. If they match, it scores 1; if one labels it as `Tie', it scores 0.5; for all other scenarios, the score is 0. Among all the annotations, we find the expert-expert score is 72.5\%. Despite the subjectivity involved in classifying responses, a reasonable level of agreement is maintained among the annotators.



\subsection{\model vs. LLaMA response}
\label{sec:investLM vs. LLaMA}
In our expert evaluation, we do not compare \model with LLaMA because LLaMA is a foundation model that is not instruction tuned, and thus we believe such a comparison would be unfair. As a case study, we explore the difference between \model and LLaMA generated response. 

Take the following financial news and instructions as an example. \textbf{Input}: \textit{U.S. investors now bake in about a 60\% chance that the Fed will raise interest rates at least another quarter point by the end of the year, pricing in relatively high probabilities of a move at the November meeting. The likelihood of a rate pause in September, however, remains near 80\%. The Fed last paused in June before raising rates in July.}. \textbf{Instruction}: \textit{Based on the news, should I buy some stocks today?}

\textbf{LLaMA's response}: \textit{"No! It is not time to buy yet. We are still waiting for the market to go down more and then we can start buying again.
 The S\&P50000 closed yesterday was -194.372 points or -0.37\%, Dow Jones Industrial Average dropped 131058points-0.7\%, Nasdaq Composite lost 10.8\% and Russell 20000000 fell 1.2.1\%."}

\textbf{\model's response}: \textit{Based on the news, the answer is not clear. While investors are anticipating more rate hikes, they are also expecting a high probability of a pause in September. The uncertainty around the Fed's next moves makes it difficult to predict how stocks will react in the near-term. There are a few factors that could influence stock prices in the coming months, including inflation levels, geopolitical events, and earnings reports. Without knowing more about your individual situation and risk tolerance, it is impossible to say if investing in stocks today is a good idea. However, it is important to do your research and understand the risks involved before making any investment decisions.}

It is evident that LLaMA's response is riddled with hallucinations, introducing numbers that are not even mentioned in the news. In contrast, \model's response is grounded in the information presented in the news and reflects logical reasoning with consideration of risks. This underscores the value of domain instruction tuning.


\section{How does \model generalize to financial NLP tasks?}
We further evaluate \model's performance on financial NLP benchmarks, following the evaluation protocol used in \citep{shah2022flue}. For the financial sentiment analysis task, we consider Financial PhraseBank (FPB) dataset \citep{malo2014good}, FOMC dataset \citep{shah2023trillion}, FiQA dataset \citep{FiQASA}, and FinSent \citep{huang2023finbert,yang2020finbert}. For the financial document classification task, we consider ESG, FLS, and QA datasets. For the financial numerical reasoning task, we consider the FinQA dataset \citep{chen2021finqa}. For the financial text summarization task, we consider the ECTSum dataset \citep{mukherjee2022ectsum}. Among the datasets, \textbf{FinSent}, \textbf{ESG}, \textbf{FLS} and \textbf{QA} are proprietary datasets that we developed for financial NLP model evaluation \footnote{\textbf{FinSent} is a sentiment classification dataset containing 10,000 manually annotated sentences from analyst reports of S\&P 500 firms. \textbf{ESG} is a text classification dataset that evaluates an organization’s strategy on environmental, social, and corporate governance. It contains 2,000 manually annotated sentences from firms’ ESG reports and annual reports. \textbf{FLS} is a text classification dataset that infers a firm's beliefs and opinions about it future events or results in the forward-looking statements. It contains 3,500 manually annotated sentences from the 
Management Discussion and Analysis section of annual reports of Russell 3000 firms. \textbf{QA} is a text classification dataset containing question-answering pairs extracted from earnings conference call transcripts. The goal of the dataset is to identify whether the answer is  valid to the question.}. Other datasets, including \textbf{FPB}, \textbf{FOMC}, \textbf{FiQA}, \textbf{FinQA} and \textbf{ECTSum} are publicly available. 

We consider the following LLMs, including two instruction tuned models \textbf{GPT-3.5}, \textbf{GPT-4} from OpenAI,  two financial LLMs, \textbf{BloombergGPT} (a 50B foundation model) and \textbf{FinMA} (an instruction-tuned model on LLaMA-7B), and one foundation model \textbf{LLaMA-65B}, upon which \model is built. We do not consider the FinGPT model, as it was only built for sentiment analysis at the time of writing.

\subsection{Generalizability on financial NLP tasks}
The results are presented in Table \ref{tab:financia_benchmarks}.  First, comparing \model against LLaMA-65, we can see that domain instruction tuning is very effective. In 8 out of 9 tasks, \model outperforms LLaMA-65. 
We want to emphasize that \model is not explicitly tuned on these financial NLP tasks - only a small portion of instructions are formatted using financial NLP tasks. This is very encouraging given that our domain instruction dataset only contains around 1,300 examples. Second, GPT-4 achieves the best performance in 6 out of the 9 tasks, while \model achieves the best performance in 2 out of the 9 tasks, suggesting that GPT-4 is the state-of-the-art commercial LLM.

\begin{table*}[]
\tiny
\resizebox{\textwidth}{!}{%
\begin{tabular}{lccccccc}
\hline
Dataset   & Metric       & LLaMA-65B & InvestLM & GPT-3.5 & BloombergGPT &FinMA & GPT-4  \\ \hline
FinSent   & Micro-F1     & 0.71      & 0.79     & 0.75    & -            & \underline{0.80}  & \textbf{0.81}  \\
FPB       & Micro-F1     & 0.38      & 0.71     & 0.75    & 0.51         & \underline{0.88}  & \textbf{0.90}  \\
FOMC      & Micro-F1     & 0.53      & \underline{0.61}     & 0.60    & -   & 0.52  & \textbf{0.73}   \\
FiQA & Micro-F1     & 0.75      & \underline{0.90}     & 0.77    & 0.75      & 0.87  & \textbf{0.92}    \\ \hline
ESG       & Micro-F1     & \underline{0.67}      & \textbf{0.80}   & 0.64    & -     & 0.51  & 0.63    \\
FLS       & Micro-F1     & \textbf{0.60}      & 0.51     & \underline{0.57}    & -            & 0.27  & \underline{0.57}    \\
QA        & Micro-F1     & 0.73      & \textbf{0.81}     & 0.71    & -            & 0.68  & \underline{0.78}     \\ \hline
FinQA     & Acc          & 0.23      & 0.29     & \underline{0.47}    & -            & 0.15  & \textbf{0.54}   \\ \hline
\multirow{4}{*}{ECTSum}  
          & Rouge-1      & 0.14      & \underline{0.26}     & 0.21    & -            & 0.08  & \textbf{0.30}  \\
          & Rouge-2      & 0.12      & 0.12     & \underline{0.13}    & -            & 0.01  & \textbf{0.15}   \\
          & Rouge-L      & 0.13      & \underline{0.17}     & 0.15    & -            & 0.06  & \textbf{0.20}  \\
          & CHRF++       & 23.65     & \underline{31.53}    & 29.79   & -            & 6.34  & \textbf{36.31}   \\ \hline
\end{tabular}%
}
\caption{Different LLMs performance on financial NLP benchmarks. \textbf{Bold} indicates the best performance metric, and \underline{underline} indicates the second-best performance metric.}
\label{tab:financia_benchmarks}
\end{table*}

\subsection{The benefit of domain instruction tuning}
To assess the advantages of domain instruction tuning across foundation models of varying sizes, we train an \model-7B model on the LLaMA-7B foundation model using our domain instruction dataset. The results are presented in Table \ref{tab:instruction}. Notably, the relative improvement brought about by domain instruction tuning is considerably more pronounced for the smaller 7B model compared to the 65B model. Specifically, for the LLaMA-7B model, domain instruction tuning improves performance by an average of 138.4\% across tasks. In contrast, for the LLaMA-65B model, there's a performance increment of 28.2\%.
The results indicate that in scenarios where computational constraints prevent deploying a 65B model, domain instruction tuning proves vital in optimizing the performance of the smaller model.


\begin{table}[]
\large
\resizebox{\columnwidth}{!}{%
\begin{tabular}{llp{1.5cm}p{1.5cm}p{1.5cm}p{1.5cm}p{1.5cm}p{1.5cm}}
\hline
Dataset   & Metric       & LLaMA-7B & InvestLM-7B & $\Delta$-7B & LLaMA-65B & InvestLM-65B  & $\Delta$-65B\\ \hline
FinSent   & Micro-F1     & 0.53     & 0.69        & 31.1\%      & 0.71      &  0.79         & 11.8\% \\
FPB       & Micro-F1     & 0.12     &0.74         & 523.7\%     & 0.38      &  0.71         & 89.3\% \\
FOMC      & Micro-F1     & 0.25     &0.40         & 57.6\%      & 0.53      &  0.61         & 15.6\% \\
FiQA      & Micro-F1     & 0.31     & 0.76        & 145.0\%     & 0.75      &  0.90         & 20.8\% \\ \hline
ESG       & Micro-F1     & 0.19     & 0.61        & 218.4\%     & 0.67      &  0.80         & 20.1\%\\
FLS       & Micro-F1     & 0.34     & 0.53        & 54.2\%      & 0.60      &  0.51         & -14.5\%\\
QA        & Micro-F1     & 0.72     & 0.84        & 16.5\%      & 0.40      &  0.81         & 10.8\%\\\hline
FinQA     & Acc          & 0.07     & 0.07        & -4.8\%      & 0.03      &  0.29         & 25.0\%\\ \hline
\multirow{4}{*}{EctSUM}    
          & Rouge-1      & 0.06     & 0.24        & 282.2\%     & 0.14      & 0.26          & 85.0\%\\
          & Rouge-2      & 0.06     & 0.10        & 62.9\%      & 0.05      & 0.12          & 4.4\%\\
          & Rouge-L      & 0.06     & 0.15        & 147.8\%     & 0.09      & 0.17          & 37.4\%\\
          & CHRF++       & 12.90    & 29.18       & 126.2\%     & 19.48     &31.53          & 33.3\%\\ \hline

          Avg. &&&&\textbf{138.4\%}&&&	\textbf{28.2\%}\\
          \hline
\end{tabular}%
}
\caption{Performance of LLaMA and \model of different model size. $\Delta$ column indicates relative improvement between \model and corresponding LLaMA.}
\label{tab:instruction}

\end{table}

\subsection{Do generic instructions help?}
\model is specifically tuned using domain specific instructions related to financial investment. In this section, we aim to explore whether the inclusion of general-purpose instructions, such as the instruction-following data employed in the Alpaca model \citep{alpaca}, can further enhance the model's performance in domain NLP tasks. Given that the general-purpose instruction dataset encompasses instructions related to numerical reasoning and sentiment, there is potential that integrating general-purpose instructions could also improve the model's capability in financial NLP tasks. 

To answer this question, we incorporate the instruction-following data used in the fine-tuning of the Alpaca model,\footnote{\url{https://github.com/tatsu-lab/stanford_alpaca}}, comprising 52K instructions, into our domain instruction dataset. Using this augmented dataset, we train an \model-7B+Alpaca-Instructions model. We then evaluate the utility of generic instructions on the financial NLP benchmarks, with the results detailed in Table Table \ref{tab:alpaca_instructions}. 

The results lead to an interesting finding that the inclusion of generic instructions appears to negatively impact the model's generalizability on domain-specific NLP tasks. When comparing \model-7B+Alpaca-Instructions (trained on the combined instruction dataset) to \model-7B (trained solely on the domain instruction dataset), it's evident that \model-7B consistently outperforms \model-7B+Alpaca-Instructions across all tasks.

This underscores the value of our carefully curated domain instructions. This finding suggests that rather than generating a large volume of general-purpose instructions,  creating a set of high-quality, domain-specific instructions can be more effective in tapping into a model's capabilities for domain tasks.

\begin{table}[]
\resizebox{\columnwidth}{!}{%
\begin{tabular}{lccc>{\centering\arraybackslash}p{3.5cm}}
\hline
Dataset   & Metric       & LLaMA-7B & InvestLM-7B &  InvestLM-7B+Alpaca-Instructions \\ \hline
FinSent   & Micro-F1     & 0.53     & \textbf{0.69 }       & 0.64   \\
FPB       & Micro-F1     & 0.12     &\textbf{0.74}        & 0.42  \\
FOMC      & Micro-F1     & 0.25     & \textbf{0.40}        & 0.32  \\
FIQA ABSA & Micro-F1     & 0.31    & \textbf{0.76}         & 0.40     \\\hline
ESG       & Micro-F1     & 0.19    & \textbf{0.61}         & 0.48   \\
FLS       & Micro-F1     & 0.34    & \textbf{0.53}         & 0.17   \\
QA        & Micro-F1     & 0.72    & \textbf{0.84}         & 0.40     \\\hline
FinQA     & Acc  & 0.07    & \textbf{0.07}         & 0.03  \\ \hline
\multirow{4}{*}{EctSUM}    
          & Rouge-1      & 0.06     & \textbf{0.24 }       & 0.14  \\
          & Rouge-2      & 0.06    & \textbf{0.10}         & 0.05  \\
          & Rouge-L      & 0.06    & \textbf{0.15}         & 0.09  \\
          & Bert Score   & 0.73    & \textbf{0.78}         & 0.75  \\
          & CHRF++       & 12.90   & \textbf{29.18}        & 19.48 \\ \hline
\end{tabular}%
}
\caption{Performance of \model-7B trained using different instruction dataset.}
\label{tab:alpaca_instructions}
\end{table}

\section{Related Work}
\noindent\textbf{Financial Domain LLMs.} Both industry and academia are closely examining the use of LLMs to analyze vast volumes of financial text, aiming to enhance investment decision-making and risk management capabilities. Before the release of ChatGPT, several financial domain language models, such as FinBERT \citep{huang2023finbert}, were developed based on BERT's architecture. Inspired by the remarkable performance of ChatGPT, efforts have been made to build financial domain large language models. Among these, BloombergGPT stands out as the first foundation model with 50B parameters, trained using Bloomberg's internal corpora \citep{wu2023bloomberggpt}.  However, BloombergGPT is not publicly accessible.  FinMA tunes LLaMA using publicly available NLP benchmarks formatted as instructions \citep{xie2023pixiu}. \model differs from BloombergGPT in that we use instruction tuning, whereas, to our knowledge, BloombergGPT serves as a foundation model. \model also diverges from FinMA, while we instruction-tune LLaMA using a carefully curated dataset that spans a wide range of financial and investment topics, FinMA relies on publicly available NLP benchmarks formatted as instructions, resulting in a more limited instruction coverage. 
Similarly, FinGPT is a LLM fine-tuned on the News and Tweets sentiment analysis dataset \citep{yang2023fingpt}, which likely limits its generalizability to other financial NLP tasks. To our knowledge, \model is the first open-source financial domain LLM that can provide insightful responses to investment-related questions, as corroborated by financial professionals.

\noindent\textbf{Instruction Tuning.}  Prior work has shown that instruction tuning — finetuning large language models on a collection of NLP tasks formatted with instructions — can enhance the language model's capability to perform an unseen task from an instruction \citep{wei2021finetuned,sanh2021multitask}. Several general-purpose instruction datasets have since been developed, including Natural Instructions \citep{mishra2021cross}, the Flan collection \citep{longpre2023flan}, the OIG dataset \citep{OIG}, among others. Recent studies have proposed automated methods for constructing instruction datasets \citep{wang2022self}. Using various instruction datasets for training on LLaMA, a series of LLaMA-based LLMs have been developed, including Alpaca \citep{alpaca} and Vicuna \citep{chiang2023vicuna}. However, most of these instruction datasets are tailored for training general-purpose LLMs. Motivated by the less-is-more-for-alignment (LIMA) \citep{zhou2023lima}, which posits that limited instruction tuning data is sufficient to guide models towards generating high-quality output, we carefully assemble a financial domain instruction dataset, specifically designed to provide helpful responses for investors and financial professionals.

\section{Conclusion}
In this work, we present a new financial domain LLM, \model, instruction tuned on the LLaMA-65B foundation model, using a set of manually crafted instruction datasets covering diverse financial and investment related topics. \model shows strong capabilities in understanding financial text and offers helpful insights in response to investment-related inquiries. Expert rates \model's responses as comparable to those of state-of-the-art commercial LLMs. Moreover, our work sheds light on using a small yet high-quality instruction dataset to fine-tune a large foundational model, suggesting a promising approach for crafting domain-specific LLMs.

\section*{Ethics Statement}
It is essential to recognize that the responses from \model should not be construed as definitive investment advice. All investment strategies and decisions should be made after careful consideration of one's financial situation, risk tolerance, and investment goals. We strongly advocate for potential investors to consult with professional financial advisors and consider multiple information sources before making any investment decisions. Relying solely on \model's outputs without thorough due diligence can lead to unintended financial consequences. Thorough analyses should be conducted to understand the strengths and weaknesses of \model.

\bibliography{anthology,custom}
\bibliographystyle{acl_natbib}

\appendix


\section{Appendix A: Training Format}
 The training format draws inspiration from the Stanford Alpaca model \cite{alpaca}. Given the instructions and input, predict the output.  
 
\textit{"prompt with input": "Below is an instruction that describes a task, paired with further context. Write a response that appropriately completes the request. Instruction:\{instruction\} Input:\{input\} Response:"
}

\textit{"prompt without input": "Below is an instruction that describes a task, paired with further context. Write a response that appropriately completes the request. Instruction:\{instruction\} Response:"
}
\label{sec:training_format}

\section{Appendix B: Expert Evaluation Interface and GPT-4 Evaluation Prompt}\label{sec:interface}
Figure \ref{tab:expert eval interface} presents the interface we used to collect expert preference judgments. Experts were given scenario simulations and moderate guidance to identify the appropriate response. Figure \ref{tab:gpt-4 eval prompt} shows the prompt of using GPT-4 to evaluate. Due to GPT-4 preferring not to make financial recommendations, the prompt was adjusted to ignore the risk of providing investment advice.

\begin{figure*}[htbp]
\small
\begin{tabularx}{\textwidth}{p{0.98\linewidth}}
\hline
\textcolor{ultramarine}{Please read the following news article and the question, and compare two responses generated by the AI assistant. You need to choose which response is better.}
\\ \\
\textcolor{ultramarine}{A better response does not need to explain a lot of additional information, and directly expresses the point of view which can be valuable for readers seeking quick, straightforward advice.} 
\\ \\
\textcolor{ultramarine}{Do not consider the risk associated with providing specific buy or sell suggestions.}
\\ \\
\textbf{News article}: \{Input\}  
\\ \\
\textbf{Question: \{Question\} }
\\ \\
\textbf{Response A}: \{Response A\} \qquad   \textbf{Response B}: \{Response B\}
\\ \\
\textbf{Please choose an option from the following:} \\
Response A is significantly better.\\
Response B is significantly better.\\
Neither is significantly better.\\
\\ \hline
\end{tabularx}
\caption{GPT-4 Evaluation Prompt}
\label{tab:gpt-4 eval prompt}
\end{figure*}

\begin{figure*}[htbp]
\small
\begin{tabularx}{\textwidth}{p{0.98\linewidth}}
\hline
\textcolor{ultramarine}{Please read the following news article and the question, and compare two responses generated by the AI assistant. You need to choose which response is better.}
\\ \\
\textcolor{purple}{Note that a better response is more accurate, concise and logical, and can be helpful for your investment decision making.} 

\\ \\
\textbf{News article}: \{Input\}  
\\ \\
\textbf{Question: \{Question\} }
\\ \\
\textbf{Response A}: \{Response A\} \qquad   \textbf{Response B}: \{Response B\}
\\ \\
\textbf{Decide which answer offers better financial investment advice.} \\
Response A is significantly better.\\
Response B is significantly better.\\
Neither is significantly better.\\
\\ \hline
\end{tabularx}
\caption{Expert Evaluation Interface}
\label{tab:expert eval interface}
\end{figure*}

\end{document}